\documentstyle{article}

\input{tcilatex}

\begin{document}

\title{Cellular automaton model for immunology of tumor growth.}
\author{M. V. Voitikova \\
Institute of Molecular\&Atomic Physics, NASB, Minsk, Belarus\\
voitikova@imaph.bas-net.by}
\date{}

\begin{abstract}
The stochastic discrete space-time model of an immune response on tumor
spreading in a two-dimensional square lattice has been developed. The
immunity-tumor interactions are described at the cellular level and then
transferred into the setting of cellular automata (CA). The multistate CA
model for system, in which all states of lattice sites, composing of both
immune and tumor cells populations, are the functions of the states of the
12 nearest neighbors. The CA model incorporates the essential features of
the immunity-tumor system. Three regimes of neoplastic evolution including
metastatic tumor growth and screen effect by inactive immune cells
surrounding a tumor have been predicted.

KEYWORDS: cellular automata, immunology of tumor growth , fractals, physics
computing, physiological models.
\end{abstract}

\maketitle

\section{ Introduction}

Practically all forms in nature are products of different kind of growth.
Mathematical growth models play an important role in research and
description of dynamics of various biological populations, which can
demonstrate both chaotic and/or regular behavior. Interest in mathematical
modeling in immunology of tumor growth arises from the ability of models to
describe and predict neoplastic growth and immune response on this growth.
In our paper, we have investigated the growth model with kinetic mechanism
for the immune response (T- lymphocyte) on tumor spreading in 2D-space. The
immunity-tumor interactions have been described at the cellular level and
then transferred into the setting of cellular automata (CA) simulation. CA
\cite{Wolfram} are mathematical idealizations of biological system where
space and time are discrete, and cell states take on the finite set of
discrete values. CA exhibit a large variety of dynamical behaviors from
fixed-point convergence and periodical motion to spatio-temporal chaos.

In numerous experiments, it has been shown that cells execute random walks
in a culture, and after moving in a certain direction cells can migrate or
turn in new direction randomly or in response on some intercellular
interactions with tumor cells. We assume that immune cell can execute random
walks in free space or move towards tumor cells. After the tumor cell
division, both daughter cells occupy 2 nearest neighbor sites if they are
empty. Here we deal with CA, which can have the set of discrete space, time,
orientation of moving cell, and limited number of different parameters
indicating activity, living time, ets. Because of its inherent computational
efficiency, CA calculational method is successfully applied to predator-prey
ecosystems \cite{Gaylord1996} , interfacial diffusion fronts \cite
{Gaylord1996} \cite{Gaylord1995}, traffic flow \cite{Schadschneider} \cite
{Nashidate}, autowave splitting \cite{Munuzuri}, quantum mechanics \cite
{Meyer}, lava flows \cite{Miyamoto}, the prolifiration dynamics \cite{Lee},
the outbreeding populations \cite{Tainaka}.

The purpose of the paper is to imitate tumor spreading in 2D-space using a
CA calculational method and the rule-based programming style. We model the
rules governing the tumor-immunity interaction in order to understand and
make clear how intercellular interactions, random collision, and cell moving
affect the tumor colony growth.

The growth model presented here is more complicated than usual models in
statistical mechanics because the cell colony configuration at a given time
step is not independent from previous configurations but is correlated in a
complicated way, i.e. tumor-immunity interaction is non-Markovian process.
The computer program developed is rather suitable and its graphic
availability will be helpful in developing the understanding in the field of
immunology of tumor growth.

Let us begin with a brief definition of immune response and tumor growth on
a regular square lattice.

\section{ The initial configuration and occupation rules.}

Biological growth is determined by tumor cell division (immune cell doesnt
divide), stimulated and spontaneous immune cell influx, and intercellular
interactions. The model presented takes in to account the immune elimination
capacity, $a$, the growth rate of a tumor population with the parameter $%
\theta $ of competition for resources, the rate of immune cells
inactivation, $\beta $, as a result of the immunity-tumor interaction, the
accumulation of immune effector cells, $c$, in a region of a tumor
(immunogenic tumor stimulates an influx of immune cells), is the spontaneous
influx, $j$, of the active immune cells into the region of the tumor, and
the rate of the immune cell elimination, $\gamma $, resulting from cell
destruction or migration out of the localization area \cite{Kuznetsov-book}
\cite{Kuznetsov}. The random distribution of immune cells and the tumor cell
seed in the center of the square lattice are used to create the initial
dynamical system configuration, and then is transferred into the setting of
CA. It has been assumed that the tumor remains stationary while immune cells
execute random walks in free space or move to any of the 4 adjacent sites,
namely, north ($N$), east ($E$), south ($S$), and west ($W$) (Fig.\ref
{fig:figure1}), if that sites are empty. Each immune cell has state values
determined by the movement orientation, the time clock indicating the living
time, and the indicator of activity (active or inactive). Each tumor cell
has a state value indicating the time till division. To define the CA moving
rules in a 2D- rectangular lattice, it is necessary to consider 12 nearest
neighbor sites near cell, as shown in Fig.\ref{fig:figure1}.

\FRAME{ftbpFU}{3.3875in}{2.5581in}{0pt}{\Qcb{ Illustration of interaction
range of a cell.}}{\Qlb{fig:figure1}}{figure1.gif}{\special{language
"Scientific Word";type "GRAPHIC";maintain-aspect-ratio TRUE;display
"ICON";valid_file "F";width 3.3875in;height 2.5581in;depth
0pt;original-width 3.2413in;original-height 2.4422in;cropleft "0";croptop
"1";cropright "1";cropbottom "0";filename
'C:/RITA/preprint98/figure1.gif';file-properties "XNPEU";}}

The rules governing the behavior of the tumor-immunity system are
implemented as the rewrite rules for all sites of the cellular automata
lattice. It should be noted, that all growth rules are probabilistic, i.e.
contain a random choice. Each time step increases the integer time clock
value of the tumor or the immune cell by 1, unless the tumor cell division
or the immune cell death occurs. In this case a new tumor cell occupies any
empty nearest neighbor place, if it exists, or an immune cell is discarded
from the immune cell population leaving an empty site, respectively. When
the immune cell interacts a tumor cell occupying an adjacent site, it can
eliminate a tumor cell with the probability a and occupies this site, or the
immune cell is inactivated by a tumor cell with the probability b. In this
case the immune cell remains in its site, but belongs to the inactive cell
population. We also used the rules that prohibit more than one walker
occupying a given site at a given time step. The immune cell moving to an
adjacent empty site or to a site occupied by a tumor cell faced by an
another immune cell remains in its site if its living time is larger than
the one of an another immune cell. So, the first immune cell holds its
orientation, but increases its time. Applying these update rules
simultaneously to all the lattice sites of the system, we can simulate the
tumor-immunity interaction in a 2D-space. The tumor growth can be stopped
after n time steps or when tumor cells reach the lattice boundary.

\section{Results.}

After n growth rules applications the immunity-tumor system yields the tumor
cluster and the distributions of active and inactive immune cells. Three
regimes of the immune response and tumor dynamics have been determined.
These are:

\begin{itemize}
\item  normal wound healing (the total number of tumor cells tends to zero),

\item  uncontrolled (exponential) tumor growth,

\item  noise-like chaotic tumor-immunity behavior.
\end{itemize}

The presented growth model shows that the immune response on 2D-space tumor
spreading exhibits oscillatory fluctuations (see Fig.\ref{fig:figure2} for
the second regime), and that the normal wound healing, when the amount of
tumor cells is equal to 0 after n time steps, has a probabilistic character.
The probability of normal wound healing increases as the division time, d,
is greater then the immune cell living time, but for a considerable amount
of inactivated immune cells the increase in the living time leads to a
screening effect by inactive immune cells surrounding tumor cell.

\FRAME{ftbpFU}{4.9191in}{2.8686in}{0pt}{\Qcb{ Tumor growth dynamics (N
curve) and leukocyte dynamics (L curve).}}{\Qlb{fig:figure2}}{figure2.gif}{%
\special{language "Scientific Word";type "GRAPHIC";display "ICON";valid_file
"F";width 4.9191in;height 2.8686in;depth 0pt;original-width
5.6161in;original-height 2.7415in;cropleft "0";croptop "1";cropright
"1";cropbottom "0";filename 'C:/RITA/preprint98/figure2.gif';file-properties
"XNPEU";}}\bigskip \FRAME{ftbpFU}{2.7311in}{2.5149in}{0pt}{\Qcb{Distribution
of tumor cell and leukocytes.}}{\Qlb{fig:figure3}}{figure3.gif}{\special%
{language "Scientific Word";type "GRAPHIC";maintain-aspect-ratio
TRUE;display "ICON";valid_file "F";width 2.7311in;height 2.5149in;depth
0pt;original-width 2.6083in;original-height 2.3999in;cropleft "0";croptop
"1";cropright "1";cropbottom "0";filename
'C:/RITA/preprint98/figure3.gif';file-properties "XNPEU";}}

Figs.\ref{fig:figure2}\ref{fig:figure3} show the typical pattern of the
uncontrolled tumor spreading when the immune response failed due to the low
elimination capacity and the fast tumor growth, where $N(t)$ is the number
of tumor cells and $L(t)$ is the number of leukocytes.

Fig.\ref{fig:figure3} shows the spatial patterns of the tumor spreading at
time step 100 on the $50\times 50$ lattice. Active and inactive immune cells
are indicated in gray (inactive cell - in light gray), tumor cells are in
black, and the empty sites are in white. Starting from random immune cell
configurations and the tumor seed in the center, the immunity-tumor system
forms a cluster structure depending on the parameters of the model. The
parameters chosen are: $a=0.5,d=\gamma =4,\beta =0.3$, and the accumulation
rate of immune cells is $c$ = 0.8. The tumor growth model exhibits a new
specific type of dynamical scaling which is intimately related to the
geometrical form. Geometrical properties of this tumor colony may be
approximated as fractal \cite{Feder} and its fractal dimension has been
estimated is equal to 1.33006.

The third regime of the growth model presented shows drastic oscillatory
fluctuations and exhibit noise-like chaotic behavior (Fig.\ref{fig:figure4}).

\FRAME{ftbpFU}{300.125pt}{173.5625pt}{0pt}{\Qcb{ Number of tumor cells vs
time.}}{\Qlb{fig:figure4}}{figure4.gif}{\special{language "Scientific
Word";type "GRAPHIC";maintain-aspect-ratio TRUE;display "ICON";valid_file
"F";width 300.125pt;height 173.5625pt;depth 0pt;original-width
4.1252in;original-height 2.3748in;cropleft "0";croptop "1";cropright
"1";cropbottom "0";filename 'C:/RITA/preprint98/figure4.gif';file-properties
"XNPEU";}}

The parameters chosen are: $a=0.82$, the time of division is equal $4$, the
immune living time $\gamma $ is equal $6$, $\beta $ $=0.3$, and the
immune-cell accumulation rate $c$ is $0.08$. This regime has been analyzed
by the Hurst's approach \cite{Feder}.

\FRAME{ftbpFU}{3.0156in}{2.4171in}{0pt}{\Qcb{Hurst's exponent vs number of
observations.}}{\Qlb{fig:figure5}}{figure5.gif}{\special{language
"Scientific Word";type "GRAPHIC";maintain-aspect-ratio TRUE;display
"ICON";valid_file "F";width 3.0156in;height 2.4171in;depth
0pt;original-width 4.1667in;original-height 3.333in;cropleft "0";croptop
"1";cropright "1";cropbottom "0";filename
'C:/RITA/preprint98/figure5.gif';file-properties "XNPEU";}}

Fig.\ref{fig:figure5} shows that the time series of the number of tumor
cells $N(t)$ and also the number of leukocytes $L(t)$ follow the empirical
law, $R/S=A\tau ^{H}$ , where rescale range $R(\tau )$ is

\begin{equation}
{\bf R(}\tau )=\max \stackrel{n}{\stackunder{u=1}{\sum }}N(u)-\overline{N}%
-\min \stackrel{n}{\stackunder{u=}{\sum }}N(u)-\overline{N}
\end{equation}

where $\overline{N}$ is the mean of the series $(1\leq n\leq \tau )$, and
the standard deviation of the cell number at the interval of observation, 20
to 500, $S(\tau )$ is

\begin{equation}
S{\bf (}\tau )=\sqrt{\frac{1}{\tau }\stackunder{u=1}{\sum^{\tau }}(N(u)-%
\overline{N})^{2}}
\end{equation}

The Hurst's exponent is defined from the slope of the asymptotic line, $H$,
is $0.82$ and $A$ is $0.48$ (Fig.\ref{fig:figure5}).

As is seen, the curves $N(t)$ and $L(t)$ have the selfaffine character, and
a fractal dimension of a trajectory as a curve can be estimated, $%
D_{F}=2-H=1.18$. The determined value of $H>0.5$ indicates the strictly
persistent character of the tumor growth with low noise level. The tumor
growth persistency allows to predict statistically the amplitude and time of
the tumor colony value at the future time intervals on the basis of already
observed data.

\section{Conclusions.}

In this paper, we have developed the model that uses the cellular automata
to imitate the process of the tumor-immunity interaction and presents
simulation results carried out on 2-dimensional grid. Our model is
computationally efficient and simulates movement, growth and interaction
tumor and immune cells better than models using differential or difference
equations. The immune response on tumor spreading in $2D$-space exhibits
oscillatory fluctuations and dependence on initial distribution, so that the
normal wound healing has a probabilistic character. The probability of
normal wound healing increases as the division time is greater then the
immune cell living time, but in the case of considerable immune cell
inactivation an increase in living time leads to the effect of screening by
inactive immune cells surrounding tumor cell.

It has been found that the dynamics of the interacting populations shows
self-organization and depends of the various experimental parameters and
initial configurations. Three regimes of the immunity-tumor interactions
have been determined. The results of simulation reveal a significant
increase in the growth rates with increasing of average number of inactive
immune cells, and decreasing of the active immune cell elimination capacity.
Time series generated by the model follows the empirical Hursts law and
demonstrates the strictly persistent character of the tumor growth with low
noise level.

The computer program developed is rather suitable and its graphic
availability is helpful in investigation of tumor growth immunology and for
creation the control strategies in immune therapy.

\end{document}